\documentstyle[12pt]{article}

\newcommand{\be}{\begin{equation}}
\newcommand{\ee}{\end{equation}}

\begin{document} 

\title{\bf 
Quark--hadron duality for $K^0-\bar{K^0}$ mixing 
}

\author{
  A.A.Penin and A.A.Pivovarov\\
  {\small {\em Institute for Nuclear Research of the
  Russian Academy of Sciences,}}\\
  {\small {\em 60th October Anniversary
  Pr., 7a, Moscow 117312, Russia.}}
  }
\date{}
\maketitle

\begin{abstract}
Long distance contribution 
to the $K^0-\bar K^0$ mixing 
is taken into account consistently
and the corrections to the naive duality result 
represented by the famous box diagram 
are found to be small.
Estimates are given in the leading order of the chiral
perturbation theory.
\end{abstract}

\hspace{3mm} PASC numbers: 14.40.Aq, 11.30.Rd, 12.38.Lg

\thispagestyle{empty}

\newpage

Both experimental and theoretical investigation of the $K^0-\bar K^0$
system is still drawing much attention since it can be used for
precise tests of the standard model of particle interactions 
and for searching new physics beyond the standard model.
Mixing of neutral mesons was calculated  in \cite{GaiLee} in the
leading order of perturbation theory (PT)
having given also a successful prediction for
$c$-quark mass through GIM mechanism \cite{GIM}.  
However it has later been pointed out
that the local effective
$\Delta S=2$ Lagrangian 
cannot account for the long distance contribution
which is present in the initial Green's function for
the $K^0-\bar K^0$ mixing and is connected with the
propagation of the light $u$-quark round the loop of the box diagram
\cite{Wolf,Hill}. 
Within effective theory 
approach
transitions with $\Delta S=2$
are given by the second order of S-matrix expansion in 
the effective $\Delta S=1$ Lagrangian 
that is given by 
colorless four-quark operators.
Long distance contributions 
are then related to the fact  
that 
the corresponding
two-point correlator of four-quark operators
reveals a nonperturbative behavior at large distances due to formation
of intermediate meson states \cite{Wolf,Hill,BigSan,Don,long}. 
The estimates of the hadronic 
contribution obtained in this
way suffer from large uncertainties because 
at present the relevant
matrix elements (or form factors)
cannot be directly extracted from experimental data 
with sufficient accuracy 
(see, for example, ref. \cite{BigSan}).
Moreover a consistent separation of
long and short distance contributions is not thoroughly 
defined in the standard
approach and hadronic 
contributions are added independently to 
the result of
perturbative calculations. 
This can lead to a
double counting 
because of duality between quark and hadron description 
of mixing. Indeed, due to GIM cancelation 
the main contribution to the PT result   
comes from  $u$-quark with a momentum smaller than 
$m_c$ running inside the loop of the box diagram. 
On the other hand this 
perturbative QCD part after inclusion of
non-perturbative QCD corrections 
has to reproduce all the hadronic contribution 
below the scale $m_c$ according to QCD-hadron duality.
Furthermore, since the perturbative result essentially saturates the 
experimentally observed $K_L-K_S$ mass difference \cite{nlcorr1}
one can conclude that the PT contribution
is dual to  the hadronic one
up to small corrections.
In the present paper we estimate
corrections to this naive duality relation.

The tree level local effective  
$\Delta S=1$   Lagrangian after decoupling 
the heavy particles ($W-$boson, t and b quarks) has the form
\begin{equation}
L_{\Delta S=1}={G_F\over \sqrt2}V_{sq'}V^*_{dq}  
\bar s \gamma_\mu (1-\gamma_5) q' 
\bar q \gamma_\mu (1-\gamma_5) d , \qquad q',~q=u,~c 
\label{ds1}
\end{equation}
where $G_F$ is the Fermi constant,  $V$ is the
Cabibbo-Kobayashi-Maskawa 
matrix of quark mixing.
  
The matrix element $M$ of the $K^0-\bar K^0$ transition
is represented by the expression
\begin{eqnarray*}
_{out}\langle\bar K^0(k')|K^0(k)\rangle_{in}=i(2\pi)^4\delta (k-k')M,
\label{amp}
\end{eqnarray*}
\be
M={i\over2}\int dx\langle\bar K^0(k')|TL_{\Delta S=1}(x) L_{\Delta
S=1}(0)|K^0(k)\rangle .
\label{me}
\ee

The effective  $\Delta S=2$  Lagrangian follows from 
eqs.~(\ref{ds1}-\ref{me})
\begin{equation}
L_{\Delta S=2}={\left({{G_F \sin\theta_c
\cos\theta_c}\over{\sqrt2}}\right)}^2(L_H+L_L) .
\label{ds2}
\end{equation}
Here $\theta_c$ is the Cabibbo angle 
that parameterizes the quark mixing 
in the limit of two generations
to which we restrict our consideration in the following.  
The heavy part of the transition
$$
L_H=i\int T_H(x)dx,~~T_H=T_{cc}-T_{cu}-T_{uc},
$$
\begin{equation}
T_{qq'}=\langle\bar K^0|T\bar s \gamma_\mu (1-\gamma_5)q
\bar q' \gamma_\mu (1-\gamma_5)d(x)\, 
\bar s \gamma_\nu (1-\gamma_5) q'
\bar q\gamma_\nu (1-\gamma_5) d(0)|K^0\rangle ,
\label{heavy}
\end{equation}
contains loops with virtual heavy $c$-quark in
the intermediate state while
\begin{equation}
L_L=i\int T_L(x)dx,~~T_L=T_{uu} ,
\label{light}
\end{equation}
describes the light part of the transition.

Expression~(\ref{ds2})  is finite due to $GIM$ mechanism but $L_H$
and $L_L$ separately require  regularization because they are
ultraviolet divergent. We use dimensional regularization.
At first sight, 
an explicit cutoff in the momentum space seems to be
more relevant in the  context of GIM cancelation. However,
it violates chiral symmetry which is the basis of our
analysis of low energy dynamics of strong interactions
within chiral perturbation theory (ChPT).

The heavy part of the Lagrangian
(\ref{ds2})
with virtual $c$-quark can be computed within ordinary 
PT. It corresponds to localization at distances of order 
$1/m_c$ and the result reads  
\begin{equation}
L_H=-{m_c^2\over 4\pi^2}\left(\bar{s} \gamma_{\mu}(1-\gamma_5)d\right)^2
\label{lh}
\end{equation}
that gives the standard contribution to the matrix element~(\ref{me})
\be
M_H={\left({{G_F \sin\theta_c
\cos\theta_c}\over{\sqrt2}}\right)}^2
\left(1+{1\over N_c}\right){m_c^2\over 4 \pi^2}X B_K
\label{mh}
\ee
where $X=f_K^2 m_K^2$,
\[
\langle\bar K^0|\left(\bar{s} \gamma_{\mu}\gamma_5d\right)^2|K^0\rangle
=2\left(1+{1\over N_c}\right)B_Km_K^2f_K^2 ,
\]
and $B_K=1$ in the vacuum saturation approximation. We 
consider $u$ and $d$ quarks as massless ones.

Corrections to  this result are under control.
The leading
PT corrections due to strong interactions  
has been obtained long ago in ref.~\cite{ptcorr}. Recently the 
next-to-leading perturbative 
calculation has been completed \cite{nlcorr1,nlcorr2}.
The hadronic matrix element of the four-quark
operator~(\ref{lh}) has been intensively studied during several 
last years with different
techniques \cite{bkcorr} and the final result is consistent
with the vacuum saturation hypothesis within a $20\%$ accuracy.
Regular corrections in $1/m_c$ {\it i.e.}
the contribution of 
operators
with 
dimension eight in mass units  
have also been taken into account and
happened to be small \cite{mccorr}.

Turning to 
the light part of the amplitude (\ref{ds2}),
it is not represented by the PT expansion because long distance
effects
have to be taken into account.
If one assumes the validity of formal PT expansion
and neglects the nonperturbative effects of $K^0$ ($\bar K^0$)
state formation the result for this part 
vanishes in the chiral limit within dimensional regularization. 
The aim of the present paper is to
calculate the nonperturbative deviation from this result.
   
To estimate the light part of the amplitude we use the factorization
procedure, {\it i.e.} we suppose the $K^0$ and $\bar K^0$ 
states to be interpolated  by the two-quark operators
with the wave function $W_\mu(k,x)$ \cite{wavef}
\be
\langle 0| \bar s(x)\gamma_\mu\gamma_5 Pexp\left(\int_0^xA_\mu dx_\mu\right) 
d(0)|K^0(k)\rangle =
i W_\mu(k,x).
\label{wf}
\ee
The wave function in eq.~(\ref{wf}) is normalized according to the equation
\[
W_\mu(k,0)= k_\mu f_K.
\label{wfn}
\]
The factorization 
approximation is used in the standard analysis of the 
box diagram contribution
and we are interested rather in the 
corrections connected with the non-perturbative behavior
of the $u$-quark propagator at large distance than 
in the violation of this hypothesis. 
Adopting this approximation 
one has a wave function of the kaon factored out
even for separated space-time points. 
The light part of the amplitude can now be written in 
the following form 
\begin{eqnarray*}
M_L={\left({{G_F \sin\theta_c
\cos\theta_c}\over{\sqrt2}}\right)}^2
(M^{(1)}_L+M_L^{(2)}).
\label{ml}
\end{eqnarray*}
Two terms  in the right hand side of this  equation 
are the  direct and cross  parts respectively.
In the direct contribution a pair of $\bar s$ and $d$ quarks
which interpolate $K^0$ $(\bar K^0)$ state is localized
in a space-time  point and we have
\begin{equation}
M_L^{(1)}=-\left({1\over N_c}\right)^2
f_K^2 k_\mu k_\nu i\int dxe^{ikx}\langle 0| T \bar u\gamma_\mu
(1-\gamma_5) u(x) \bar
u\gamma_\nu (1-\gamma_5) u(0)|0\rangle .
\label{ml1}
\end{equation}
Using the anomaly relation
$
\partial_\mu (\bar u\gamma_\mu \gamma_5 u)  =  2w
$
where 
\[
w(x)= {1\over 16\pi^2}{\rm tr}_{c}\tilde G_{\mu\nu} G_{\mu\nu},
\]
eq.~(\ref{ml1}) can be rewritten in the following form
\[
M_L^{(1)}= 4\left({1\over N_c}\right)^2
f_K^2 i\int dxe^{ikx}\langle 0| T w(x) w(0)|0\rangle . 
\]
In  the small $k$ limit an  expansion of the two-point correlator of
$w$ operators is known within the chiral perturbation theory (ChPT)
\cite{GassLeut} 
\[
i\int dxe^{ikx}\langle 0| T w(x) w(0)|0\rangle |_{k^2=m_K^2} =
\frac{1}{6}m_K^2(f_\pi^2+\tilde H_0)+O(m_K^4)
\]
where $\tilde H_0$ is the vacuum susceptibility \cite{WitVen}
\[
\tilde {H_0}|_{N_c\rightarrow\infty}= {1\over 2}f^2_\pi .
\]
We have used a mass relation 
\[
m_\eta^2=\frac{4}{3}m_K^2 
\]
which is valid for massless $u$ and $d$ quarks. 
Finally the direct contribution becomes
\be
M_L^{(1)}=\frac{2}{3} \left({1\over N_c}\right)^2X(f_\pi^2+\tilde H_0).
\label{ml1f}
\ee

In the cross contribution  $\bar s$ and $d$ quarks
which interpolate $K^0$ $(\bar K^0)$ state come 
from different space-time points and we have
\begin{equation}
M_L^{(2)}= \left({1\over N_c}\right)^2  
\left({1\over N_c}\Pi^s+2\Pi^o\right),
\label{ml2}
\end{equation}
where $\Pi^s$ and $\Pi^o$ are singlet and octet contributions 
respectively 
\[
\Pi^s=i \int dx 
\langle 0| T\bar u\gamma_\alpha (1-\gamma_5)u(x)
\bar u\gamma_\beta (1-\gamma_5)u(0)|0\rangle 
(2g_{\mu\alpha}g_{\nu\beta}-g_{\mu\nu}g_{\alpha\beta})
W^*_\mu(k,x)W_\nu(k,x),
\]
\[
\Pi^o=i \int dx 
\langle 0|T \bar u\gamma_\alpha t^a (1-\gamma_5)u(x) 
\bar u\gamma_\beta t^a (1-\gamma_5)u(0)\rangle
(2g_{\mu\alpha}g_{\nu\beta}-g_{\mu\nu}g_{\alpha\beta})
W^*_\mu(k,x)W_\nu(k,x),
\]
and $t^a$ are $SU(3)_{c}$ generators
normalized with the condition $tr(t^a t^b)=\delta^{ab}/2$. 
If the PT
expression is used for $u$-quark propagators this contribution
computed in the dimensional regularization is non-zero even 
in the chiral limit because the kaon wave function introduces
the nonperturbative scale $\sim \Lambda_{QCD}$.
Note that this contribution is suppressed by the factor
$(\Lambda_{QCD}/m_c)^2$ in comparison with the leading one
and is not taken into account in the standard analysis. 
Clearly, an estimate of the contribution of massive states to the cross 
part of the amplitude depends on the model for the $K$-meson
wave function.  
At the same time the contribution 
of the  $\pi^0$ and $\eta$
mesons to the color singlet part $\Pi^s$ in the chiral limit 
is proportional to $|W_\mu(k,0)|^2$ and is  model independent 
\be
M_L^{(2)}={1\over 8}\left({1\over N_c}\right)^3Xf_\pi^2.
\label{ml2f}
\ee
Contribution of states with nonzero masses are suppressed by the ratio
of the size of the wave function to the mass of the state and are neglected.
One can reasonably suppose that the magnitude 
of long distance effects in this channel is semiquantitatively represented
by this simple estimate. 
The advantage is also the additional suppression
of the nondirect contribution 
in $N_c$ counting (it is suppressed by the factor $1/N^3_c$)
that allows one not to be very demanding to 
the accuracy of this estimate.
In fact, we assume that the contribution of higher mass states 
can amount about fifty percent of the estimate (\ref{ml2f})
still not leading to the large uncertainty
of the total result.

Note that if the explicit cutoff in momentum space
$\mu$ is used in calculation of
the light part of the amplitude~(\ref{light}) for cutting high energy
contributions, the heavy part~(\ref{heavy}) of the full matrix element 
has to be properly modified. For $\mu = m_c$ the heavy part
vanishes and the PT result for the light part coincides with eq.~(\ref{mh}). 
On the other hand the light part of the amplitude can now be
computed by saturating eq.~(\ref{light}) by contributions of the 
hadrons
with masses smaller than $m_c$ what is the standard way to estimate
the long distance contribution. However in this case 
the light part (or long distance contribution)
is not an addition to~(\ref{mh}) 
but rather represents the whole amplitude. 
So the nonperturbative corrections in this case should be
associated with the difference between the hadronic and PT
expressions for the light part of the amplitude 
not with the hadronic contribution itself. 

We would like to stress this point. The physical result being
invariant, its representation through (and splitting into) long and short
distance contributions depends on regularization procedures used for
computation $L_H$ and $L_L$. In these circumstances 
the advantage of dimensional 
regularization for perturbative computation of the light part of the
whole amplitude becomes clear -- the light part vanishes.
The dominant contribution to the
amplitude stems from the heavy part which can be reliably
computed within PT while the 
light part contributes only because of 
some nonperturbative effects. 
These nonperturbative effects
under an assumption of factorization
are organized into a correlator of colorless currents of light quarks
taken at the scale of order $m_K^2$. 
ChPT is now used to compute the final answer
that is therefore exact. 
Numerically it is rather small. 
An additional advantage that 
makes the result more reliable is that 
the correlator is suppressed in formal
counting in $N_c$.
If, instead, the explicit cutoff in momentum space 
is used for regularization,   
hadronic contributions represent the total result.
Because the energy scale $m_c$ is rather large for light quarks
the spectral density cannot be accurately given that leads 
to large uncertainties \cite{BigSan}.
So, it seems impossible to estimate the corrections
to eq.~(\ref{mh}) in this way.  
Moreover the multihadron contributions 
generate a power dependence on the cutoff 
(see, for example, ref. \cite{Don}) and 
for $\mu\sim m_c$ there is no 
reliable way to find their numerical value.
On the other hand if the cutoff is small and chosen to be of order of
the chiral symmetry breaking scale  $\mu\sim m_K$, 
the contribution of hadronic states from momentum region $m_K<k<m_c$
is out of control because it can be described 
neither within pQCD no ChPT. 
This is the reason for considerable 
uncertainties of previous estimates of the
hadronic contribution to the mixing.
At the same time in the dimensional regularization we deal
with the well defined chiral expansion in $m_K^2/\Lambda_\chi^2\sim 0.25$
where $\Lambda_\chi\sim1~{\rm GeV}$ is the scale of ChPT.

The obtained results are applied to
calculation of the parameters of $K^0-\bar K^0$ system.
The correction to the $K_L-\bar K_S$
mass difference reads
\[
\Delta m_{LS}=\Delta m_{sd}
\left(1- {8\pi^2f_\pi^2\over 3B_K(N_c+1)N_c m_c^2}
\left(1+{\tilde H_0\over f_\pi^2}-{1\over 4N_c} \right)\right)
\sim\Delta m_{sd}\left(1-{0.03\over B_K}\right)
\]
where $\Delta m_{sd}$ is the short distance contribution
that follows from eq.~(\ref{mh}) and the correction comes from 
eqs.~(\ref{ml1f},~\ref{ml2f}). For numerical estimate 
we use $m_c=1.4~{\rm GeV}$, $f_\pi=0.132~{\rm GeV}$.

Thus we found 
that the difference between the full result 
for the matrix element of the mixing and 
the contribution of the heavy part 
computed within the dimensional regularization is
about $3\%$ of the total result. 
This difference vanishes in large $N_c$ limit.

To conclude we would like to stress again 
the main result of the paper.
We showed that the contribution of light hadrons to  
$K^0-\bar K^0$ transition amplitude 
(known as dispersive or long distance contribution as well)
should not be added independently
but is essentially dual to PT expression of the box diagram
with virtual $u$-quark. 
In our analysis we rely on factorization approximation even for
nonlocal operators and intensively use the chiral limit of QCD.
Within these assumptions duality is rather accurate 
quantitatively 
and the PT expression represents the complete answer well
while the genuine nonperturbative contribution
of light quarks to mixing is negligible (about 3\%).
It has been proven that 
the same result can be also obtained by direct summation of hadronic
contributions till energies of order $m_c$. 
The latter method is however impractical at present
because of poor experimental data.

\vspace{5mm}
\noindent
{\large \bf Acknowledgments}\\[1mm]
This work is supported in part by Russian Basic Research Foundation
grant N~97-02-17065. The research of A.Pivovarov is also
supported by Russian Basic Research Foundation
grant N~96-01-01860.

\end{document}